\documentclass[prd,reprint,superscriptaddress,nofootinbib,floatfix]{revtex4-2}
\usepackage{braket}
\usepackage{amsmath,amssymb,amsfonts,bm}
\usepackage{subfig}
\usepackage{graphicx}
\usepackage{float}
\usepackage[unicode=true,
            pdfusetitle,
            colorlinks=true,
            urlcolor=blue,
            anchorcolor=black,
            citecolor=red,
            filecolor=navyblue,
            linkcolor=red,
            menucolor=navyblue,
            linktocpage=true,
            pdfproducer=medialab]{hyperref}
\hypersetup{pdfauthor={Name}}

\begin{document}

\preprint{}

\title{Quantum Ising Model on (2+1)-Dimensional Anti–de Sitter Space\\using Tensor Networks}

\author{Abhishek Samlodia}
\email[]{asamlodi@syr.edu}
\affiliation{Syracuse University, Department of Physics, Syracuse, NY 13244, USA}

\author{Simon Catterall}
\email[]{smcatter@syr.edu}
\affiliation{Syracuse University, Department of Physics, Syracuse, NY 13244, USA}

\author{Alexander F. Kemper}
\email[]{akemper@ncsu.edu}
\affiliation{Department of Physics and Astronomy, North Carolina State University, Raleigh, NC 27695, USA}

\author{Yannick Meurice}
\email[]{yannick-meurice@uiowa.edu}
\affiliation{Department of Physics and Astronomy, University of Iowa, 30 N Dubuque St, Iowa City, IA 52242, USA}

\author{Goksu Can Toga}
\email[]{gctoga@ncsu.edu}
\affiliation{Department of Physics and Astronomy, North Carolina State University, Raleigh, NC 27695, USA}

\date{\today}

\begin{abstract}
We study the quantum Ising model on (2+1)-dimensional anti-de Sitter space using Matrix Product States (MPS) and Matrix Product Operators (MPOs). We explore the bulk phase diagram of the theory on regular tessellations of hyperbolic space with coordination number seven and find disordered and ordered phases separated by a phase transition. We find that the boundary-boundary spin correlation function exhibits power law scaling deep in the disordered phase of the Ising model consistent with holography. At the critical point, we find the boundary entanglement entropy scales logarithmically with subsystem size but away from this, we see a linear scaling. In comparison, the full system exhibits a volume law scaling, which is expected in chaotic and/or highly connected systems. We also measure Out of time Ordered Correlators (OTOCs) to explore the scrambling behavior of the theory.
\end{abstract}

\maketitle

\section{Introduction}
\label{sec:intro}

Gravitational theories are conjectured to be holographic in nature; that is the gravitational degrees of freedom associated with some region of spacetime can be represented by quantum fields residing on the boundary of that region \cite{Susskind,RevModPhys.74.825}. In the case of an anti-de Sitter spacetime, the asymptotic boundary is conformally flat, and the boundary theory is a conformal field theory \cite{Maldacena}. This holographic correspondence typically maps a weakly coupled bulk gravitational theory to a strongly coupled boundary theory. This motivates one to try to study such theories using non-perturbative methods such as Monte Carlo sampling.

Monte Carlo methods work for Euclidean studies of many such systems but sign problems appear for real-time dynamics and certain fermionic/complex-action theories. Hence, one is led to investigate numerical methods that can be used in a Hamiltonian formulation. In principle, such calculations could be done using quantum computers, but the current hardware and algorithms impose limitations on the size of the system. In (1+1) dimensions, classical tensor network methods provide extremely efficient alternatives \cite{SWhite,SCHOLLWOCK201196}. However, it is generally thought that such methods will not be effective when they are lifted to (2+1) dimensional systems \cite{SCHOLLWOCK201196} due to the need for very large bond dimensions. In this paper we will show that the constant ratio of bulk to boundary sites, which is typical of hyperbolic lattices, allows us to use MPS methods reliably in system sizes larger than a regular $2d$ square lattice. In fact these methods can be adapted to study such systems with as many as a few hundred lattice sites.

\begin{figure*}
    \centering
    \includegraphics[width=\textwidth]{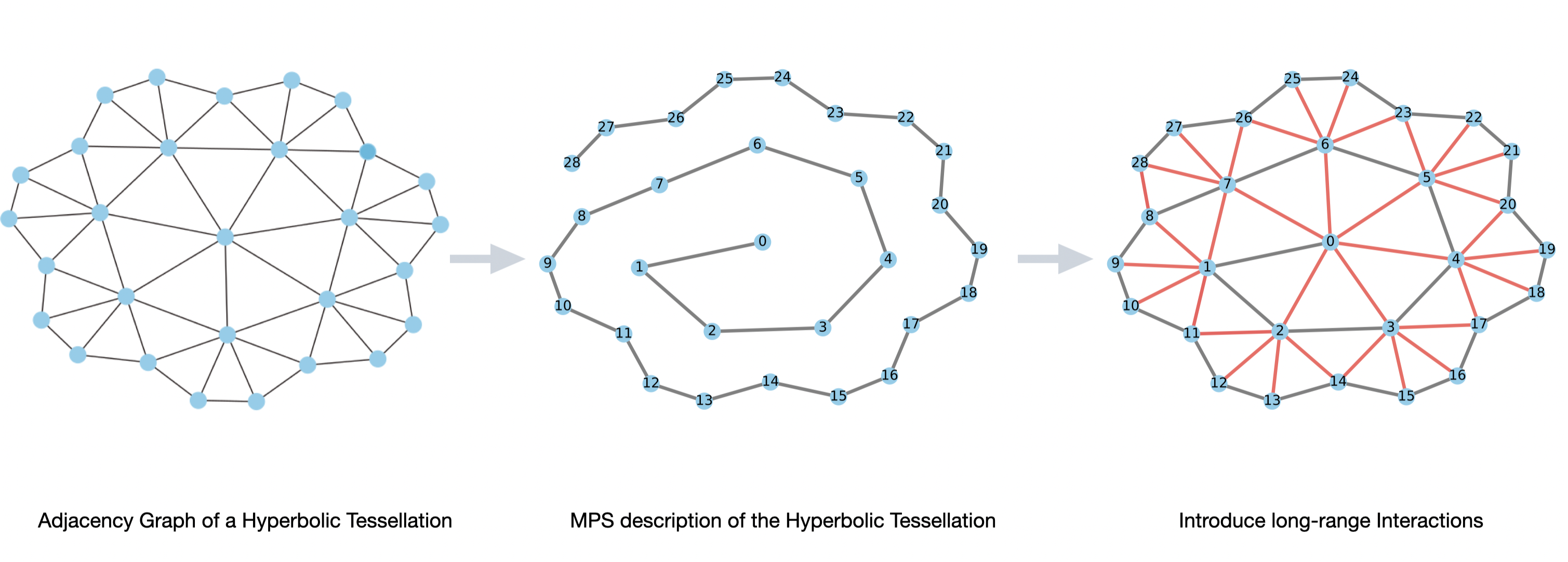}
    \caption{A summary of our MPS construction adapted to hyperbolic tessellations.}
    \label{fig:main_fig}
\end{figure*}

Specifically, we have used methods based on Matrix Product States (MPS) to study the ground state and time evolution of the quantum Ising model formulated on a discretization of two-dimensional hyperbolic space, which we can think of as a spatial slice of three-dimensional anti-de Sitter space. We investigate the phase structure of this theory and the properties of the associated boundary theory which is obtained by tracing out the bulk degrees of freedom. It should be noted from the outset that since the bulk tessellation is fixed, the construction does not include any gravitational fluctuations, so any effective boundary theory may not contain a local energy-momentum tensor. Nevertheless, the model can be seen as a step along the path toward the construction of a true discrete holographic model that can be explored using future quantum computers.  

Using tensor networks or quantum circuits to simulate holographic physics is not a new idea, and there have been a great number of earlier studies on related topics. The most well-known is the Multi-Scale Renormalization Ansatz (MERA)\cite{swingle_entanglement_2012}, which uses a set of disentangling matrices to remove local entanglement between sites thereby achieving the entanglement scaling expected for gapless theories. Furthermore, several works have implemented tensor networks using hyperbolic tessellations~\cite{Okunishi_2024, oshima2024ferromagneticisingmodelhierarchical,swingle_constructing_2012,singh_holographic_2018,dey_simulating_2024,asaduzzaman2024quantum,toga2025fast}, in addition to tessellations of superconducting qubits\cite{kollar2019hyperbolic,boettcher2020quantum} and circuit implementations of hyperbolic tessellations\cite{Li:2023vwx,lenggenhager2022simulating}. It has been shown that Matchgate Tensor network calculations, which can be related to disordered Ising models, capture holographic expectations very well~\cite{jahn_boundary_2022,jahn_holography_2018}.   

Our work, in contrast, starts with a disorder-free transverse field with a longitudinal symmetry-breaking field and uses a matrix product state (MPS) ansatz to capture the wavefunction of low-lying states.  To obtain the ground state of this system, we use the Density Matrix Renormalization Group (DMRG) algorithm \cite{white1992}. In addition to this, we also study the dynamical properties of this model by representing the Hamiltonian as a Matrix Product Operator (MPO) and implementing time evolution by using the Time Evolving Block Decimation (TEBD) algorithm \cite{schollwock2011}. Using TEBD we calculate Out-of-Time-Ordered Correlators (OTOCs)~\cite{sekino_fast_2008,Maldacena:2015waa,swingle2016measuring,xu_accessing_2020,swingle2018unscrambling,xu2024scrambling} and investigate quantum chaos and information propagation in this system.

Our goal in this paper is to see how far we can push simple tensor network methods like TEBD and DMRG  to study holographic physics and where these approaches break down, signaling the need for better algorithms or the use of quantum algorithms. The structure of the paper is as follows: we start by briefly describing MPS/MPO method's basic idea in Section~\ref{sec:mps_mpo_description}. Hamiltonian for the model is introduced in Section~\ref{sec:setup}. In Section-\ref{sec:results} we first show results for the Ising model on a square lattice in flat space and then find the critical point for this model in $AdS_3$. This is followed by studies of the two-point boundary correlation function, the entanglement entropy, and investigations of the scrambling behavior using OTOCs. Lastly, we conclude the paper in section-\ref{sec:conclusions}.

\section{Matrix Product State and Matrix Product Operator}
\label{sec:mps_mpo_description}
In general, any quantum state can be expanded on the tensor product of the Hilbert spaces associated with the quantum spins on each lattice site
\begin{equation}
    \Psi=\sum_{i_1,\ldots i_N} c_{i_1\ldots i_N}\ket{s_1, s_2\cdots s_N}
\end{equation}
The MPS ansatz assumes that the coefficients $c_{i_1\ldots i_N}$ can be
written as the trace of a product of matrices $M$ of size $D_{\rm bond}$ - the bond dimension:
\begin{equation}
    c_{i_1\ldots i_N}={\rm Tr}\,\left(M^{i_1}\ldots M^{i_N}\right)
\end{equation}
The size of $D_{\rm bond}$ controls the amount of entanglement that can be faithfully captured in the approximation. This construction is particularly well-suited for systems whose low-energy states satisfy an area law, such as those encountered in spin chains and gauge theories discretized on one-dimensional spatial lattices.
To describe operators we use a Matrix Product Operator(MPO) description as follows, 
\begin{equation}
    \hat{O} = \sum_{s,s'} {\rm Tr}(W^{s_1,s_1'}\dots W^{s_N,s_N'})\ket{s_1, s_2\cdots s_N}\bra{s^\prime_1, s^\prime_2\cdots s^\prime_N}
\end{equation}
MPOs provide a natural representation for local Hamiltonians, which we use in our DMRG and TEBD studies. An MPO can be applied to an MPS in the following way:




\begin{align}
\hat{O}\ket{\psi} 
&= \sum_{\vec{s},\,\vec{s}'} 
   \Big(
       W^{s_1,s_1'} 
       W^{s_2,s_2'} 
       \cdots
   \Big)
   \Big(
       M^{s_1'} 
       M^{s_2'} 
       \cdots
   \Big)
   \ket{\vec{s}} 
   \notag\\[2mm]
&= \sum_{\vec{s},\,\vec{s}'} 
   \sum_{\vec{\alpha},\,\vec{\beta}}
   \Big(
       W^{s_1,s_1'}_{\beta_1,\beta_2} \,
       W^{s_2,s_2'}_{\beta_2,\beta_3} \,
       \cdots
   \Big)
   \Big(
       M^{s_1'}_{\alpha_1,\alpha_2} \,
       M^{s_2'}_{\alpha_2,\alpha_3} \,
       \cdots
   \Big)
   \ket{\vec{s}}
   \notag\\[2mm]
&= \sum_{\vec{s}} 
   \sum_{\vec{\alpha},\,\vec{\beta}}
   \Big(
       \sum_{s_1'} 
           W^{s_1,s_1'}_{\beta_1,\beta_2} \,
           M^{s_1'}_{\alpha_1,\alpha_2}
   \Big)
   \Big(
       \sum_{s_2'} 
           W^{s_2,s_2'}_{\beta_2,\beta_3} \,
           M^{s_2'}_{\alpha_2,\alpha_3}
   \Big)
   \cdots
   \ket{\vec{s}}
   \notag\\[2mm]
&= \sum_{\vec{s}} 
   N^{s_1}_{(\alpha_1\beta_1),(\alpha_2\beta_2)} \; 
   N^{s_2}_{(\alpha_2\beta_2),(\alpha_3\beta_3)} \;
   \cdots
   \ket{\vec{s}}
\label{eq:mpo_on_mps}
\end{align}

where $\ket{\vec{s}} := \ket{s_1, s_2,\ldots, s_N}$ and we have defined

\begin{equation}
    N^{s_i}_{(\alpha_i\beta_i),(\alpha_{i+1}\beta_{i+1})} \; := \; \sum_{s_i'}\, W^{s_i,\,s_i'}_{\beta_i,\,\beta_{i+1}}\; M^{s_i'}_{\alpha_i,\,\alpha_{i+1}}\,.
\label{eq:N_tensor}
\end{equation}

This is again an MPS of the same form, but with an enlarged bond dimension $D_{\mathrm{bond}} \times D_{\mathrm{MPO}}$, where
$D_{\mathrm{MPO}}$ is the bond dimension of the MPO.

Using these two building blocks, we can then obtain the ground state of a quantum system represented by an MPS using the DMRG algorithm, which uses a variational method on the matrix parameters to find an approximation to the ground state. To capture dynamics one can represent the time evolution operator as an MPO and implement discrete Trotter time evolution of the MPS.

\section{Ising model on anti-de Sitter space}
\label{sec:setup}
The Hamiltonian we study is given by,
\begin{equation}
    H_{\rm Ising} = -J_{zz}\sum_{\langle j, k \rangle} Z_j Z_k - \sum_j J_x X_j - m\sum_j Z_j
\end{equation}
where $X,Y,Z$ represent standard Pauli matrices $\sigma^x, \sigma^y, \sigma^z$ respectively and $\{j,k\}$ are lattice site indices. $J_{zz}$ is the nearest neighbor coupling applied on sites that are connected on the tessellation, and $J_x$ is the transverse coupling. We add a small mass (symmetry breaking field) $m$ to break the model's $Z_2$ symmetry which is required to lift the degeneracy in the ground state of the system.

We study this model using Hamiltonian-based, quantum-inspired tensor networks - Matrix Product States (MPS) - which give an approximate representation of the quantum state of a many-body system. As is well known, the MPS ansatz is inherently a one-dimensional construction, but it is possible to generalize the method by building an MPS that traverses through a two-dimensional lattice. This approach performs well for simulating strip geometries, but it is limited due to the scaling of the entanglement entropy in the extra dimension, which requires very large bond dimensions as the extra dimension size grows larger. 

Our objective in this work is to find out if this approach can be used to simulate models defined on hyperbolic tessellations, in which a large proportion of spins live on the boundary of the lattice, which limits the amount of long range interaction needed to capture the $2d$ dynamics. 

\begin{figure}
    \centering
    \includegraphics[scale=0.55]{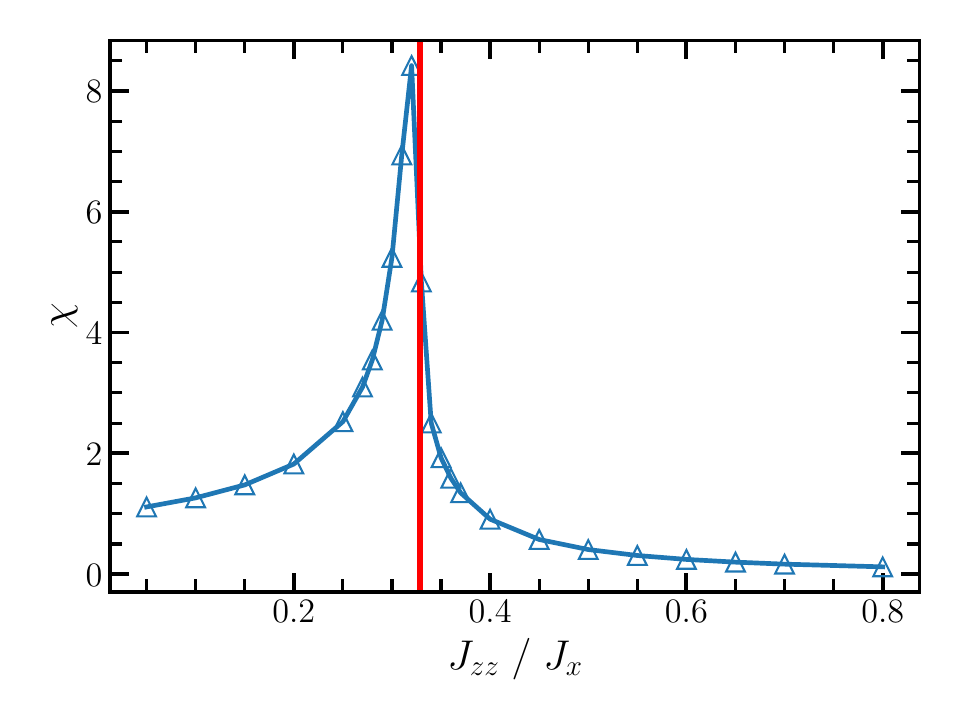}
    \caption{Bulk phase transition for 12x12 square lattice flat space Ising model with $m = 0.001$, the red line is the critical point ($J_{zz}/J_x = 0.3285$) quoted in Ref.~\cite{PhysRevLett.101.250602}}
    \label{fig:flat_ising}.
\end{figure}

We start from a tessellation of two-dimensional hyperbolic space $H^2$ and obtain a one-dimensional MPS chain by winding out from the center of the tessellation along successive layers until the boundary is reached.  The order of the matrices in the MPS ansatz follows the order of the sites as shown in Fig.~\ref{fig:main_fig} for a $(3,7)$ triangulated lattice consisting of $29$ vertices ($3$-layers) where the edges represent the $2$ body interactions and spins are located at the vertices. Grey edges represent the original MPS description and the red edges represent the long-range interactions needed to capture the $2d$ characteristics.

\section{Results}
\label{sec:results}
In this section we show results obtained using the DMRG algorithm implemented within the ITensor module~\cite{Itensor}.
\subsection{Bulk properties}
As a sanity check of our implementation of $2d$ MPS method, we first simulate the flat two-dimensional Ising model on a  square lattice, whose bulk phase transition can be captured by the magnetic susceptibility given by 
\begin{equation}
\label{eq:mag_sus}
    \chi = \frac{1}{N_{\rm sites}}\left(\bigg\langle \sum_{j,k}Z_j Z_k \bigg\rangle - \bigg\langle \sum_j Z_j \bigg\rangle^2\right)
\end{equation}
This is shown in Fig.~\ref{fig:flat_ising}. The observed critical point from the simulations is at $J_{zz}/J_x \sim 0.32(1)$ which is in good agreement with the value obtained by previous tensor network studies in two dimensions ($J_{zz}/J_x = 0.3285$) \cite{PhysRevLett.101.250602}  confirming that our implementation~\footnote{For a square lattice, the MPS chain starts from the bottom left corner of the lattice and proceeds towards the nearest neighbor on the right until last site is reached in the first row of the square grid. This last site is connected to first site of the second row of the grid and then usual right neighbor hopping is carried out until the last site of the last row at the top of the grid. Then the vertical nearest neighbor bonds are introduced followed by bonds representing appropriate boundary conditions.} can capture $2d$ physics.

\begin{figure}
    \centering
    \includegraphics[scale=0.55]{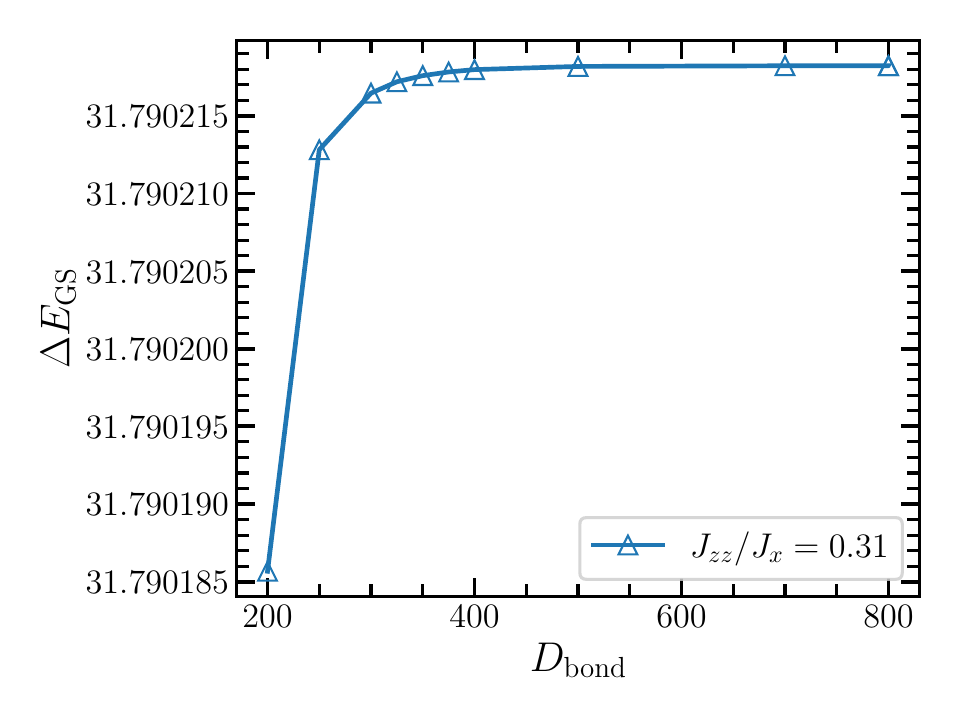}
    \caption{Dependence of energy of the ground state with bond-dimension $\chi$ for $J_{zz}/J_x = 0.31$ for three layer lattice}
    \label{fig:convergence_0p31}
\end{figure}

Having checked that our implementation works for a square lattice one can check the convergence of the ground state energy as a function of bond dimension for the case of a hyperbolic spatial lattice. Fig.~\ref{fig:convergence_0p31} shows the ground state energy obtained using the DMRG algorithm as we increase the bond-dimension $D_{\rm bond}$ at coupling $J_{zz}/J_x = 0.31$ using a three-layer tessellation.  We see that the ground state energy converges quickly with increasing bond dimension and levels off at bond dimensions higher than $400$.

\begin{figure}
    \centering
    \includegraphics[scale=0.55]{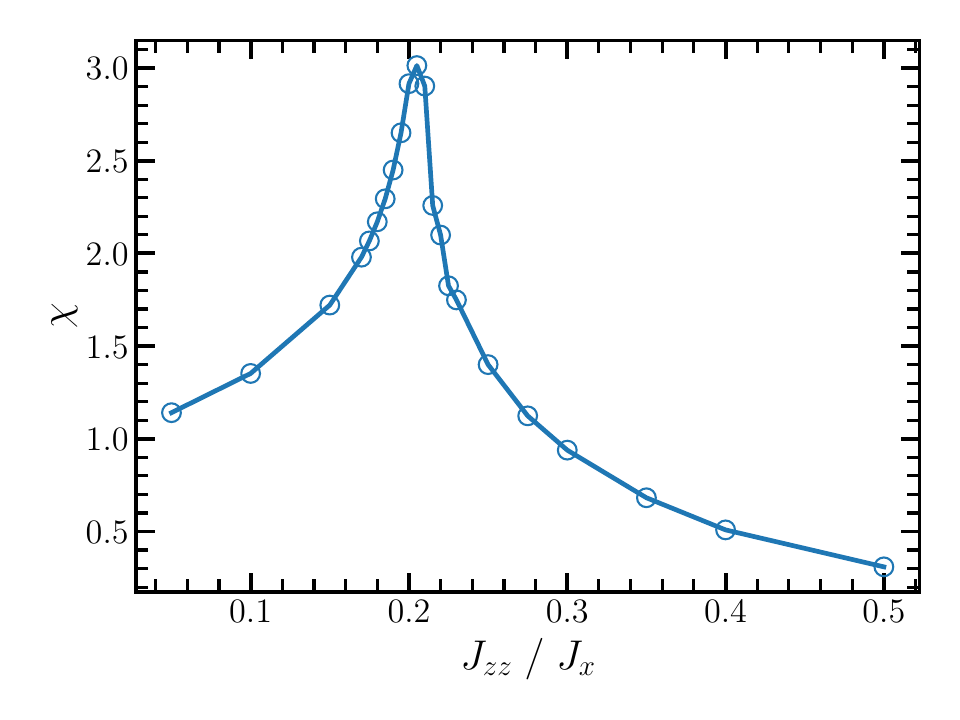}
    \caption{Bulk phase transition for Ising model on five-layer tesselated $AdS_3$ with $m = 0.001$. }
    \label{fig:bpt_ising}
\end{figure}

\begin{figure}
    \centering
    \includegraphics[scale=0.55]{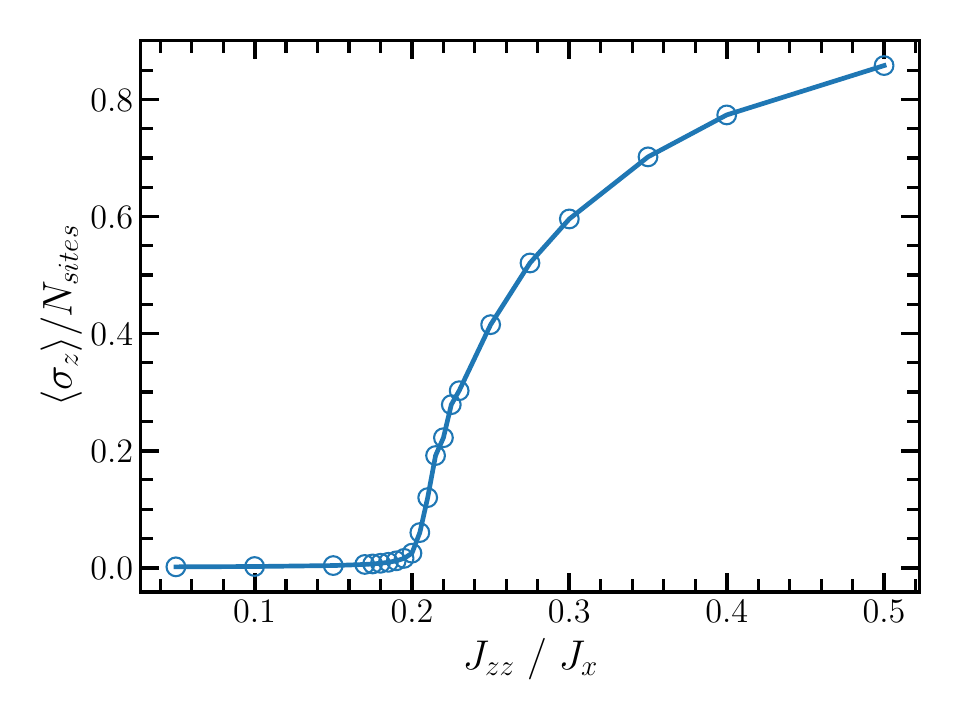}
    \caption{Magnetization for Ising model on five-layer tesselated $AdS_3$ with $m = 0.001$.}
    \label{fig:bpt_sz_ising}
\end{figure}

Similar to the flat lattice, to capture the phase transition, we investigate the magnetic susceptibility and magnetization in the hyperbolic case. In Fig.~\ref{fig:bpt_ising} and Fig.~\ref{fig:bpt_sz_ising}, we plot the magnetic susceptibility and magnetization, respectively, as a function  of 
coupling for the hyperbolic lattice model of five-layer tessellation.
Right of this transition point, we have the ordered phase, and to the left of it we have the disordered phase. 

Table~\ref{tab:lat_vol_table} lists the number of points for three different lattice volumes used in this work. The last column shows the number of boundary points in a given tessellation. 
\begin{table}[]
    \centering
    \begin{tabular}{|c|c|c|}
        \hline
         \# of layers &  \# of vertices & \# of boundary vertices\\
         \hline
         3 & 29 & 21\\
         4 & 85 & 54\\
         5 & 232 & 147\\
         \hline
    \end{tabular}
    \caption{Lattice volume for (3,7) hyperbolic tessellation}
    \label{tab:lat_vol_table}
\end{table}

\subsection{Boundary Correlators}
\label{ssec:boundary_correlators}
It is interesting to study the spin correlation function when both source and sink points are
restricted to the boundary of the tessellated disk $\delta D$. In particular,
we can plot this as a function of distance between source and
sink {\it along the boundary}. The correlation function is defined as,
\begin{equation}
    C(r) := \frac{1}{N_b} \sum_{j,k\in \delta D}\delta(r-\lvert j - k\rvert)\left(\langle Z_j Z_k \rangle - \langle Z_j \rangle\langle Z_k \rangle\right)
\end{equation}
where $N_b$ is the number of boundary vertices in the triangulation and $r$ is the boundary distance between the source and sink. 

This has been studied before using
classical simulation of scalars and Ising spins in hyperbolic space \cite{Asaduzzaman:2020hjl,Asaduzzaman:2021bcw}.
In particular, one expects the correlator to be determined by the minimal
length lattice path between source and sink, which, for the hyperbolic space, runs
through the bulk. Because of the structure of the tessellation, this minimal
path length scales only logarithmically with the boundary separation. This in turn, has the effect of producing a power law dependence on boundary distance even
for a theory that is gapped in the bulk.

To analyze the correlators, we first fold the data about $r = (N_b-1)/2$ points and then use the data smoothing procedure as explained in~\cite{Samlodia:KD_on_H2} with a smoothing window of three lattice spacings. Then we fit a power law function $f(r) = Ar^{-B} + D$ to this folded
and smoothed data. 
\begin{figure}
    \centering
    \includegraphics[scale=0.55]{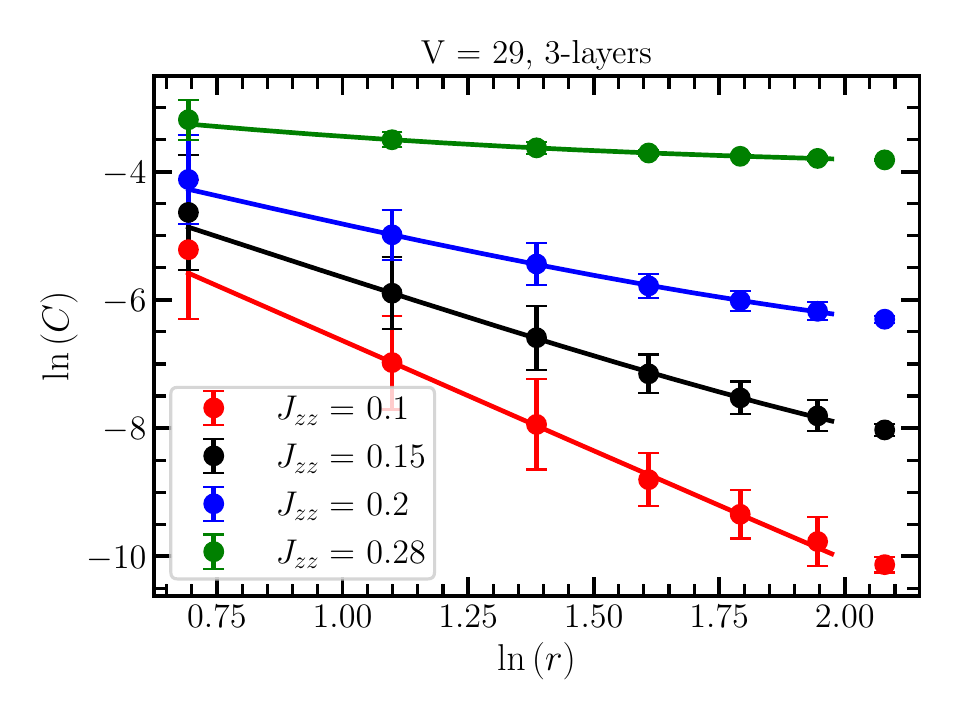}
    \caption{Boundary correlators for a 3-layer hyperbolic lattice with $m = 0.001$. Solid lines represent best fit lines, and points represent data from simulations that have been smoothed out as mentioned in the text. $R^2 \approx 0.999$ for all the fits.}
    \label{fig:V29_corr}
\end{figure}
\begin{figure}
    \centering
    \includegraphics[scale=0.55]{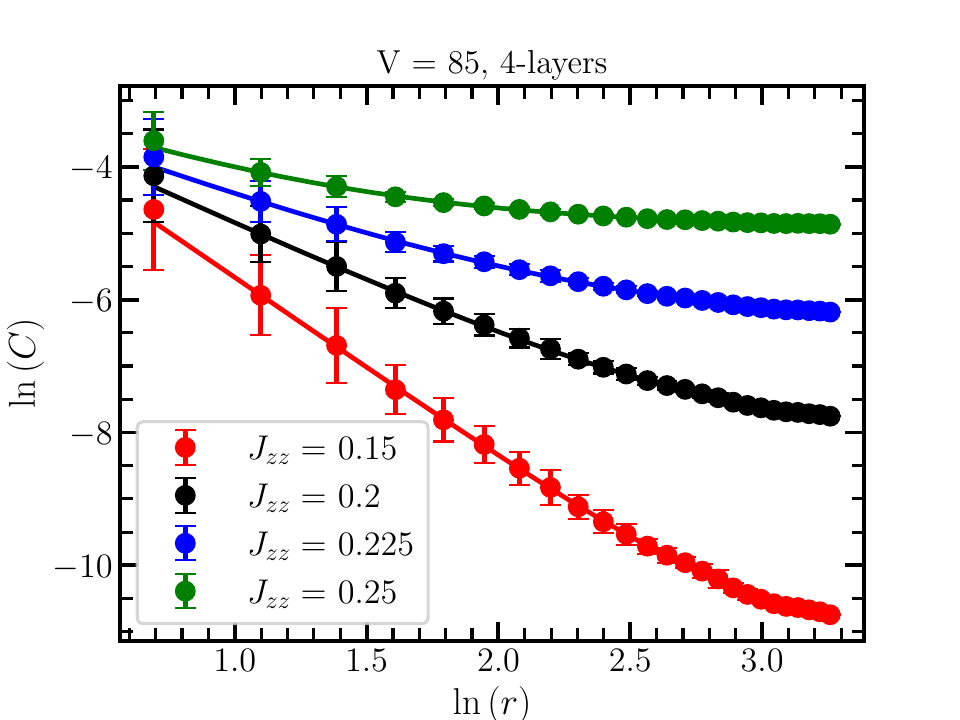}
    \caption{Boundary correlators for a 4-layer hyperbolic lattice with $m = 0.001$. Solid lines represent best fit lines, and points represent data from simulations that have been smoothed out as mentioned in the text. $R^2 \approx 0.997$ for all the fits.}
    \label{fig:V85_corr}
\end{figure}
\begin{figure}
    \centering
    \includegraphics[scale=0.55]{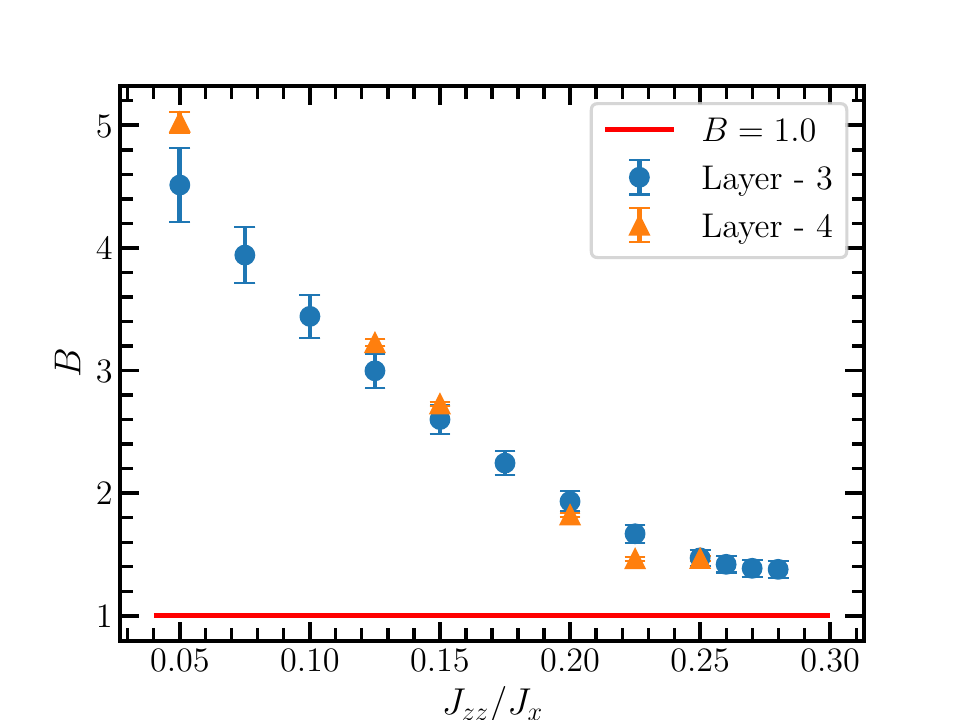}
    \caption{Variation of power $B$, obtained from fits with $m = 0.001$}
    \label{fig:power}
\end{figure}
Fig.~\ref{fig:V29_corr} and Fig.~\ref{fig:V85_corr} shows these correlator fits for the three and four layer lattice. Fig.~\ref{fig:power} shows how the power obtained from these fits varies as a function of the bulk parameter, $J_{zz}/J_{x}$ at $m=0.001$. Our fits confirm that the boundary-boundary correlation function decays as a power law, as expected. 

Notice that the power approaches a minimum as the bulk coupling is tuned towards
its critical value. For reference, we draw on the plot a horizontal line
corresponding to that expected for a two-dimensional free fermion. 

\subsection{Sub-region entropies}
In this subsection, we will study the scaling of the entanglement entropy $S_\ell=-{\rm Tr}\,\left[\rho(\ell)\log{(\rho(\ell))}\right]$ with respect to the subsystem size $\ell$. In general, to calculate the entanglement entropy, the system with $L$ spins is divided into two subsystems $A$ and
$B$ of size $\ell$ and $L-\ell$ respectively and the resultant density matrix is defined by
\begin{equation}
    \rho(\ell):={\rm Tr}_B\ket{\Psi}\bra{\Psi}
\end{equation}

Initially, let us focus just on the effective boundary theory
which can be obtained by constructing the density matrix from the MPS corresponding to the ground state of the full system and then tracing over the bulk indices. This requires partial traces over the full density matrix of the spins on the hyperbolic graph. To do this, we make use of the routines provided in the  ITensorEntropyTools.jl module~\cite{entropytools}. 

Once the reduced density matrix corresponding to the boundary is obtained, we divide the boundary
into two regions to investigate the scaling of the von Neumann entropy of the boundary theory with respect to the subsystem size. 

\begin{figure}
    \centering
    \includegraphics[scale=0.55]{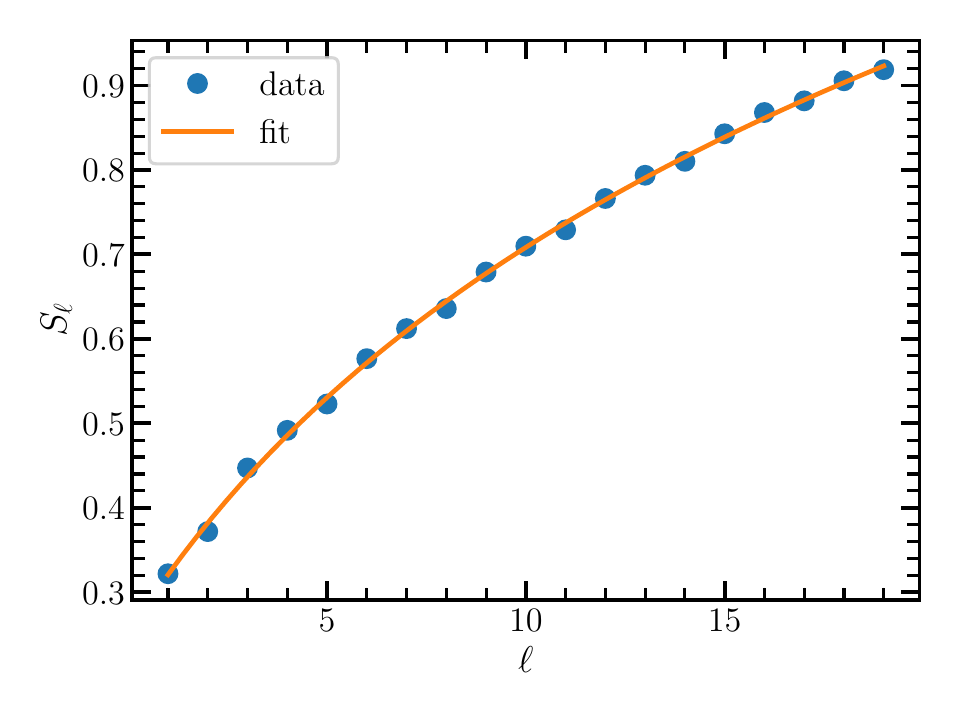}
    \caption{Variation of entanglement entropy with subregion length for $J_{zz}/J_x = 0.32$ and 3-layer lattice. The fit function is $(c/3)\log_{2}(x/a + b)$ where $c = 1.030(49)$, $a = 3.97(43)$ and $b = 1.658(25)$. $R^2 \approx 0.999$ for the fit.}
    \label{fig:boundary_entropy_0.32}
\end{figure}

\begin{figure}
    \centering
    \includegraphics[scale=0.55]{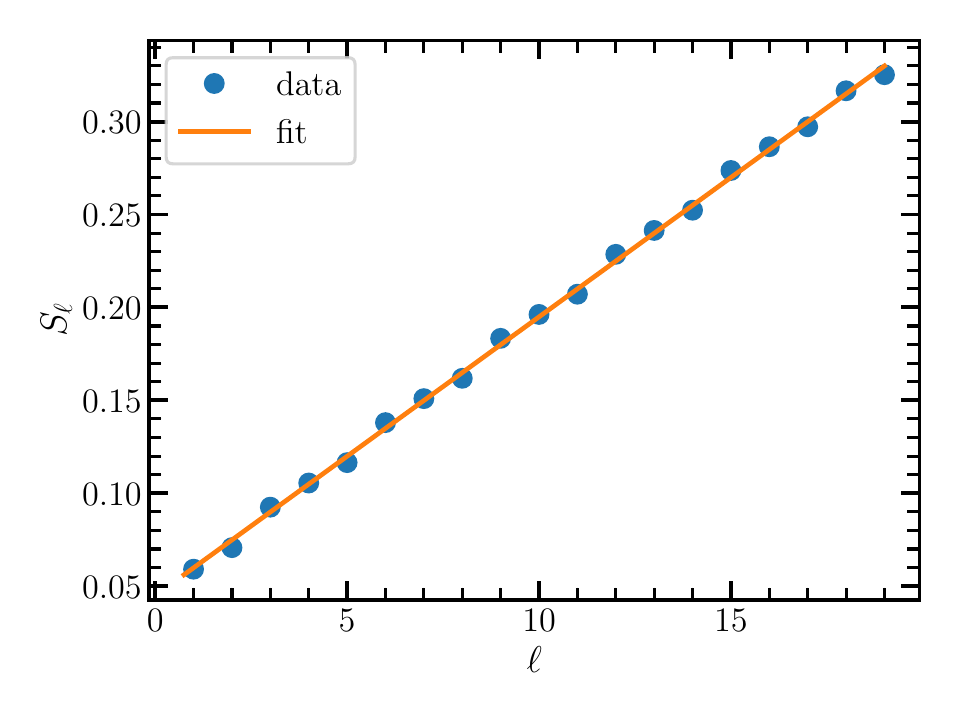}
    \caption{Variation of entanglement entropy with subregion length for $J_{zz}/J_x = 0.15$ and 3-layer lattice. The fit function is $f(x) = ax +b$ where $a = 0.01500(12)$ and $b = 0.0449(14)$}
    \label{fig:boundary_entropy_0.15}
\end{figure}

We observe that the boundary entanglement entropy scales logarithmically with subregion length at the critical point (see Fig.~\ref{fig:boundary_entropy_0.32}). 
The fitted central charge (obtained by a simple least squares
procedure) is
consistent with unity, which might be expected if the boundary CFT at the bulk critical point corresponded to a free Dirac fermion.
Away from the critical point and deeper in the disordered phase, we see a linear behavior (see Fig.~\ref{fig:boundary_entropy_0.15}). 

From our boundary correlator results, one would have expected that the boundary theory would be governed by a CFT even when the bulk is gapped. However, our measurement of the
entanglement scaling shows  no evidence of this, which we interpret as arising
from the fact that one expects the boundary theory to be governed by a non-local
Hamiltonian for a generic
value of the bulk coupling. This shows the drawback of this method in simulating AdS/CFT physics compared to MERA and similar constructions, where a critical boundary theory is built iteratively, and a logarithm is observed for all
couplings. It is possible that the desired logarithmic scaling will show up at bigger volumes with much larger bond dimension, but with our current computational resources, we see no evidence of this. 

\begin{figure}
    \centering
    \includegraphics[scale=0.55]{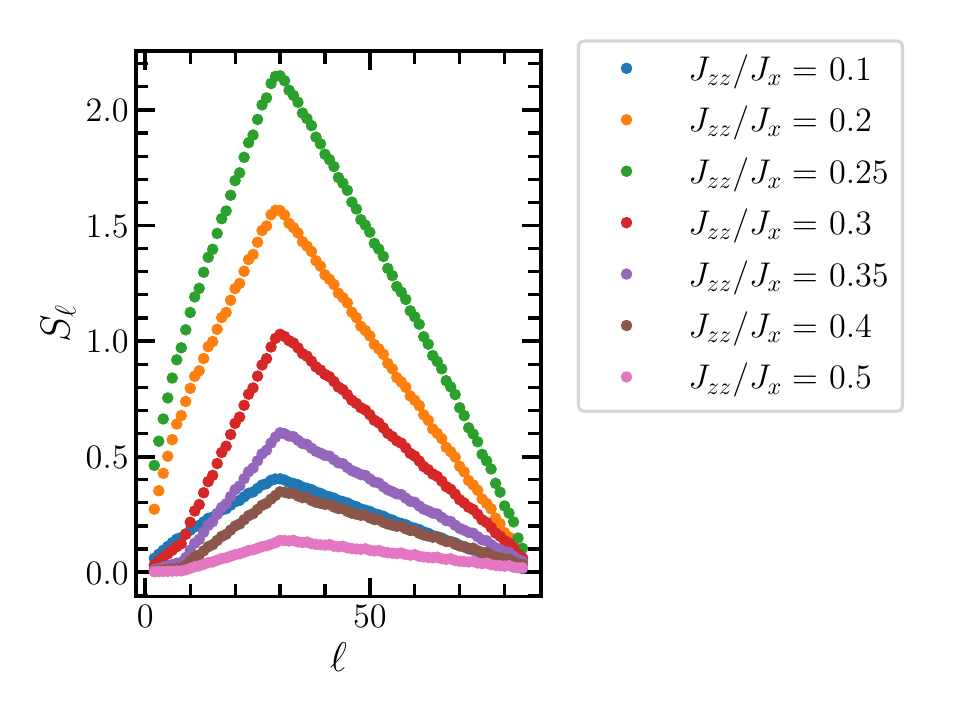}
    \caption{Variation of bulk entanglement entropy with subregion length for various $J_{zz}/J_x$ values and a 4-layer lattice.}
    \label{fig:full_entropy_nlay4}
\end{figure}

\begin{figure}
    \centering
    \includegraphics[scale=0.55]{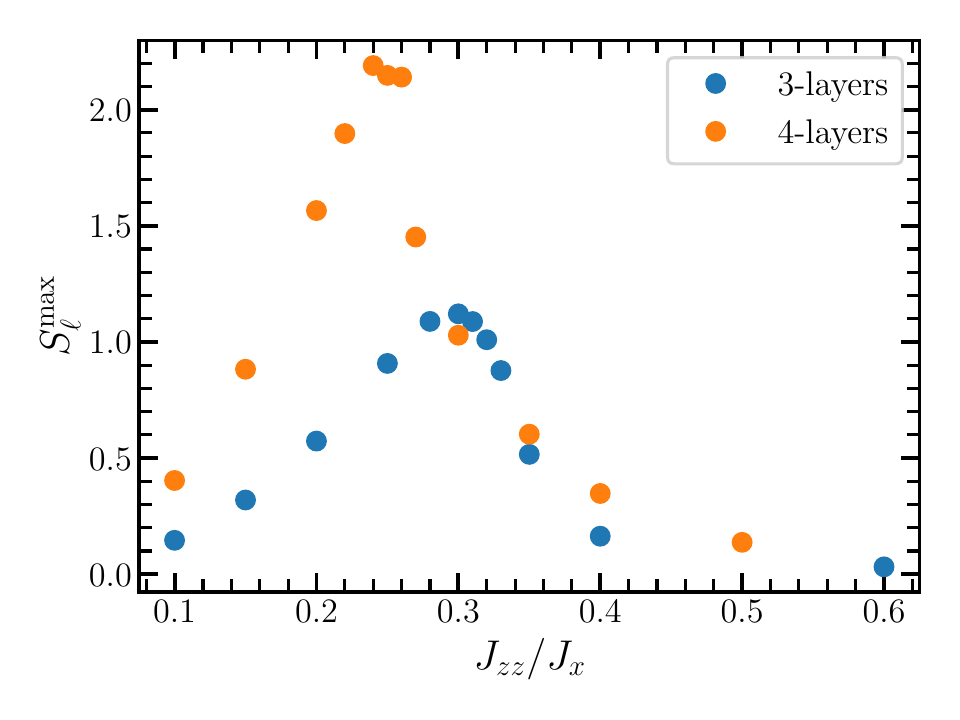}
    \caption{Variation of maximal bulk entanglement entropy with bulk coupling $J_{zz}/J_x$ for 3 and 4 layer lattice.}
    \label{fig:slmax}
\end{figure}

We can also compute the entanglement entropy for the full system  by progressively tracing
out subregions starting from the central node.
In this case, we find that the entropy peaks when the subregion
reaches the first boundary node and then decreases. This is true both at criticality and away from it in the disordered as well as ordered phase.  Fig.~\ref{fig:full_entropy_nlay4} shows these findings for a four-layer lattice, respectively. This kind of behavior is known as volume law scaling and is expected to appear in highly connected and chaotic systems~\cite{eisert2010colloquium}. Such an increasing-peak-decreasing behavior can be explained from a geometric perspective, where one can see a linear increase and then a linear decline in the number of links that cut the boundary of the lattice. Another interesting observable  which could potentially be used to locate bulk phase transitions is the behavior of the peak of the entanglement entropy in Fig.~\ref{fig:full_entropy_nlay4} as a function of bulk coupling $J_{zz}/J_x$ as shown in Fig.~\ref{fig:slmax}. This $S_\ell^{\rm max}$ which is the maximum bulk-boundary bipartite entropy peaks at the bulk critical point.

Interestingly, MPS methods that normally exhibit area law scaling capture this volume law scaling behavior when overlaid on a hyperbolic tessellation, which shows us that even though our description fails to capture the criticality of the boundary theory accurately, it still manages to capture the bulk behavior well.

\subsection{Out of time ordered correlators(OTOCs)}
In this section, we calculate out-of-time-ordered correlators, which are defined as
\begin{align}
 O_{ij}(t) & :=\langle \lvert \lvert [Z_i(t), Z_j(0)] \rvert\rvert^2\rangle\\
 & = 2\left(1-Re[\langle Z_i^\dagger(t)Z_j(0)^\dagger Z_i(t)Z_j(0)\rangle]\right)   
\end{align}
where, lattice site $i$ is the source point and $j$ is the sink. We work in the infinite temperature limit where density matrix is an identity operator.
We obtain the time evolved operators $Z_i(t)$ by expressing the Hamiltonian as an MPO 
and implementing Heisenberg evolution via the time-evolving block decimation (TEBD) algorithm. The system’s state is represented as an MPS, and time evolution under the Hamiltonian $H$ is implemented using a Trotter–Suzuki decomposition of the evolution operator $e^{-iHt}$ into a sequence of two-site unitary gates. After each local gate application, singular value decomposition (SVD) is used to truncate the small Schmidt coefficients, keeping the MPS within a fixed bond dimension~\cite{Vidal2003, Vidal2004, Daley2004, SCHOLLWOCK201196}.

These time-dependent correlators probe how the information in an initial
local quantum state spreads throughout the system as a result of
the dynamics. Generically, local quantum information is scrambled into non-local degrees of freedom under time evolution. It is closely related to the problem of thermalization
in quantum systems and to the presence of chaotic dynamics. In the latter case quantum information spreads exponentially fast through a system and
allows one to define a Lyapunov exponent from the behavior of the OTOC in a manner similar to a classical chaotic system. Holographic systems corresponding to
black holes are thought to furnish examples of so-called fast scramblers, and so it is interesting to look for such behavior in our toy holographic model.

\begin{figure}
    \centering
    \includegraphics[scale=0.55]{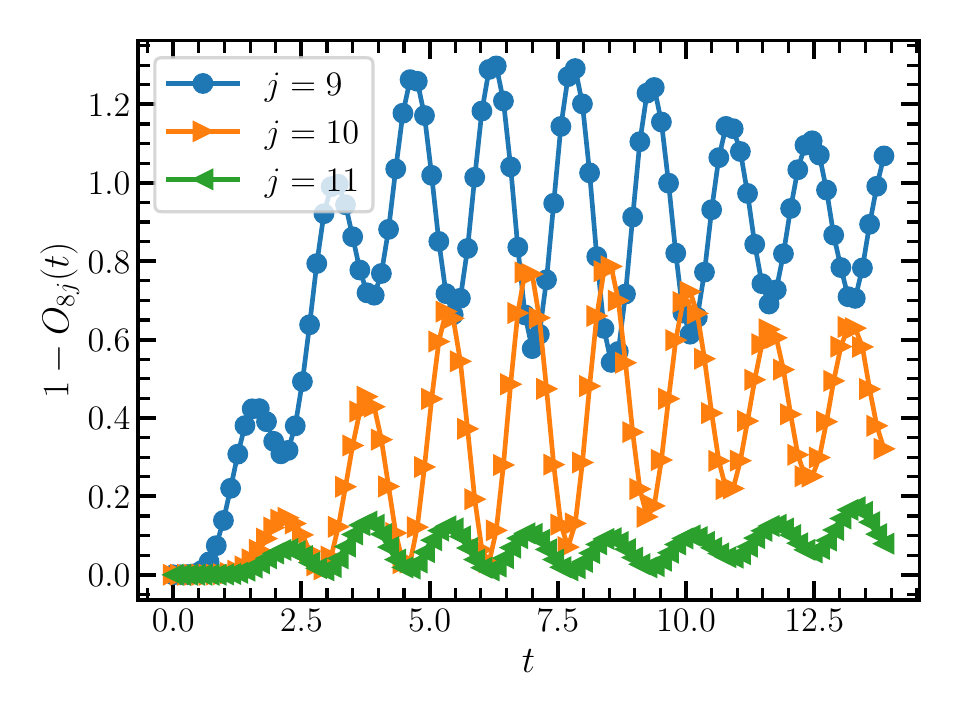}
    \caption{OTOC with source at first boundary node $i=8$ and sink at nodes on the boundary of the lattice, at $J_{zz}/J_x = 0.3$ for 3-layer lattice}
    \label{fig:otocs_3}
\end{figure}

\begin{figure}
    \centering
    \includegraphics[scale=0.55]{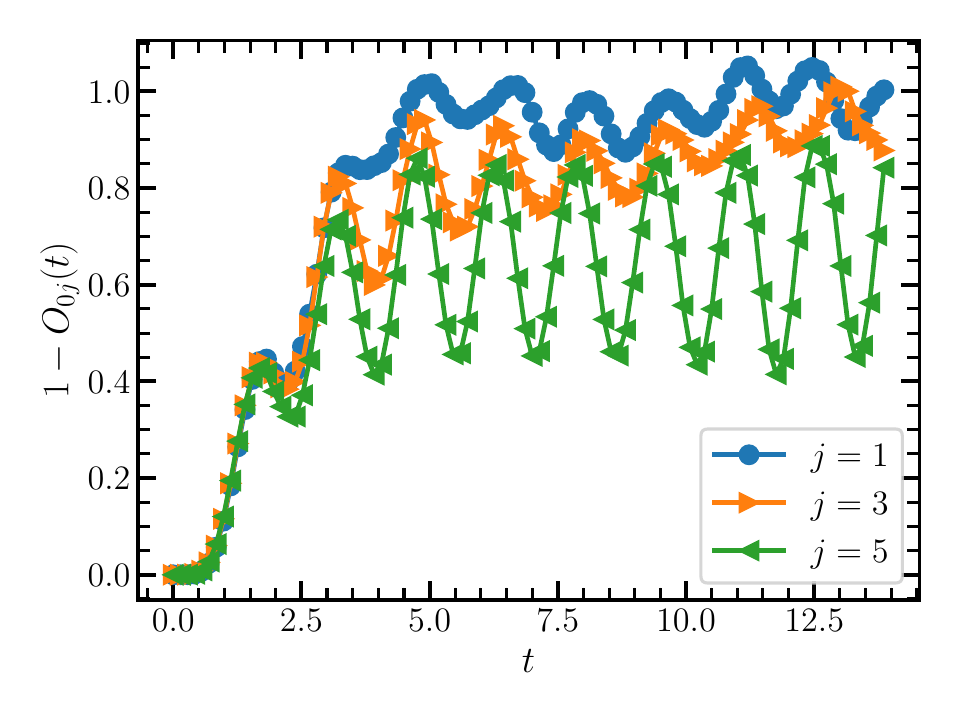}
    \caption{OTOC with source at center node $i=0$ and sink at nodes in the bulk of the lattice, at $J_{zz}/J_x = 0.3$ for 3-layer lattice}
    \label{fig:otocs_6}
\end{figure}

Our results for the time dependence of OTOCs initialized at both the center and a boundary site are given in Fig.~\ref{fig:otocs_3} and Fig.~\ref{fig:otocs_6}, respectively showing the initial rapid exponential
growth. The generic form of these corresponds to a rapid rise and then 
an oscillating approach to an equilibrium plateau.

\begin{figure*}[htbp]
    \centering
    \subfloat[$t = 0.98$, source at center node]{
        \includegraphics[width=0.3\textwidth]{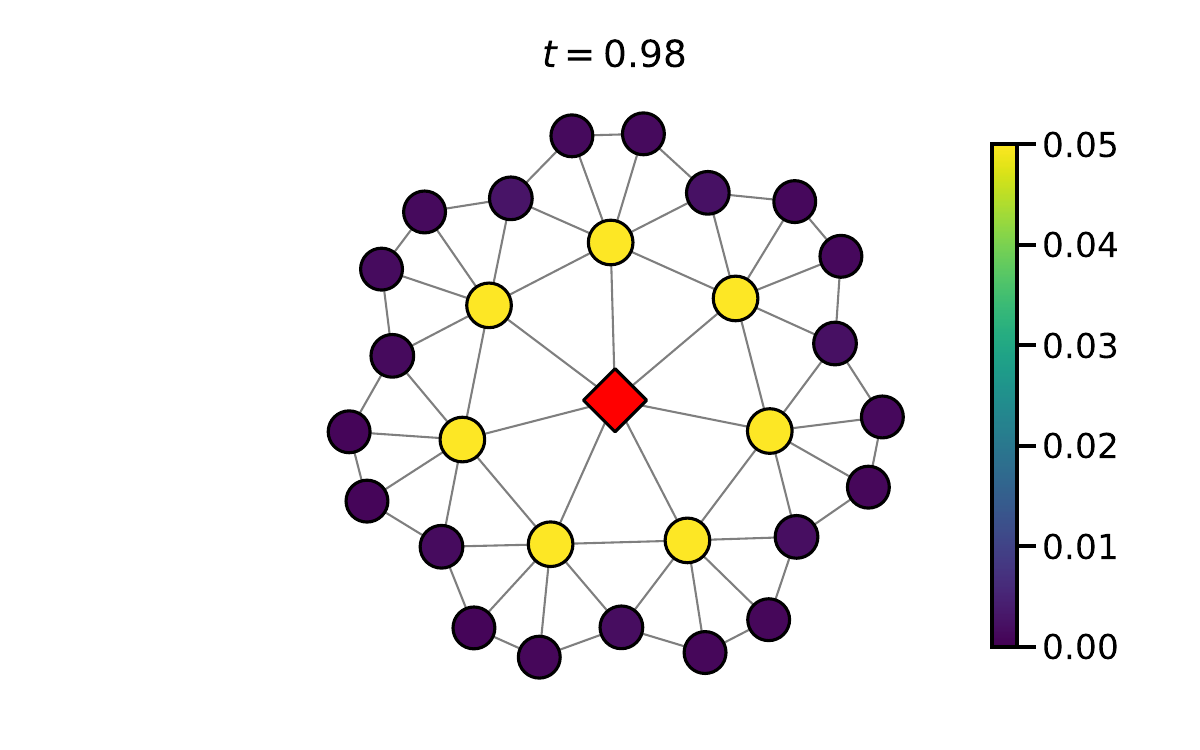}
    }\hfill
    \subfloat[$t = 1.54$, source at center node]{
        \includegraphics[width=0.3\textwidth]{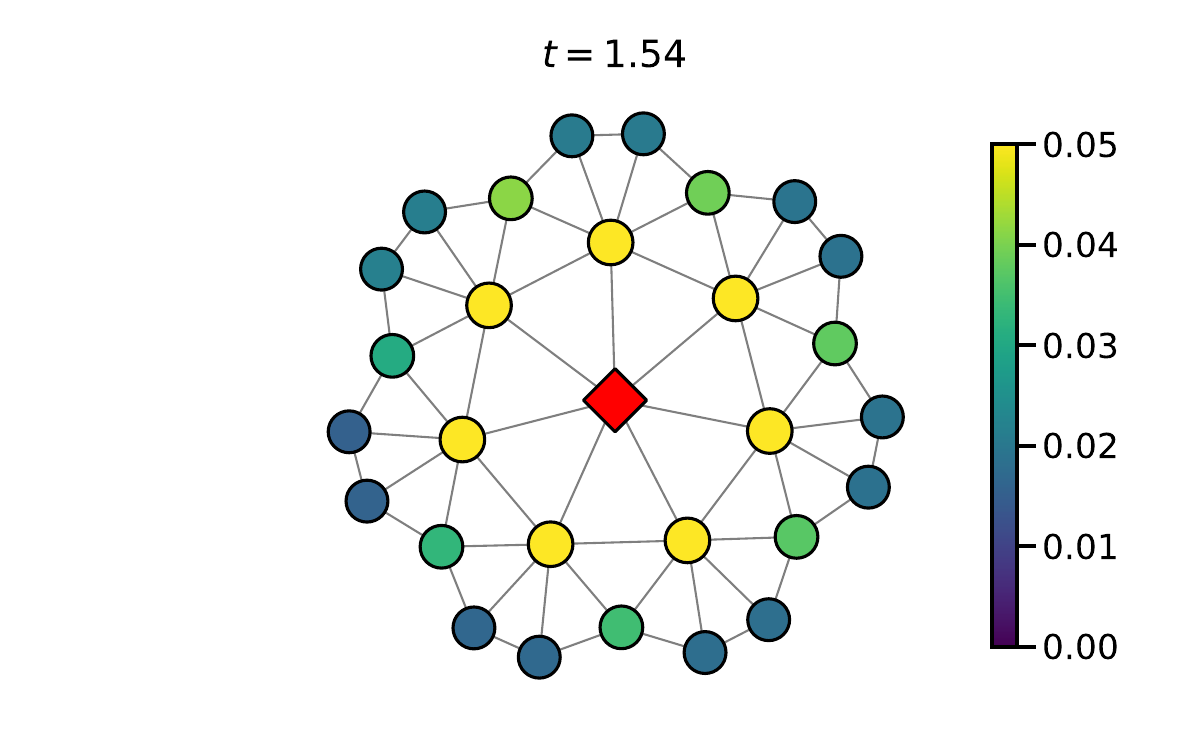}
    }\hfill
    \subfloat[$t = 2.1$, source at center node]{
        \includegraphics[width=0.3\textwidth]{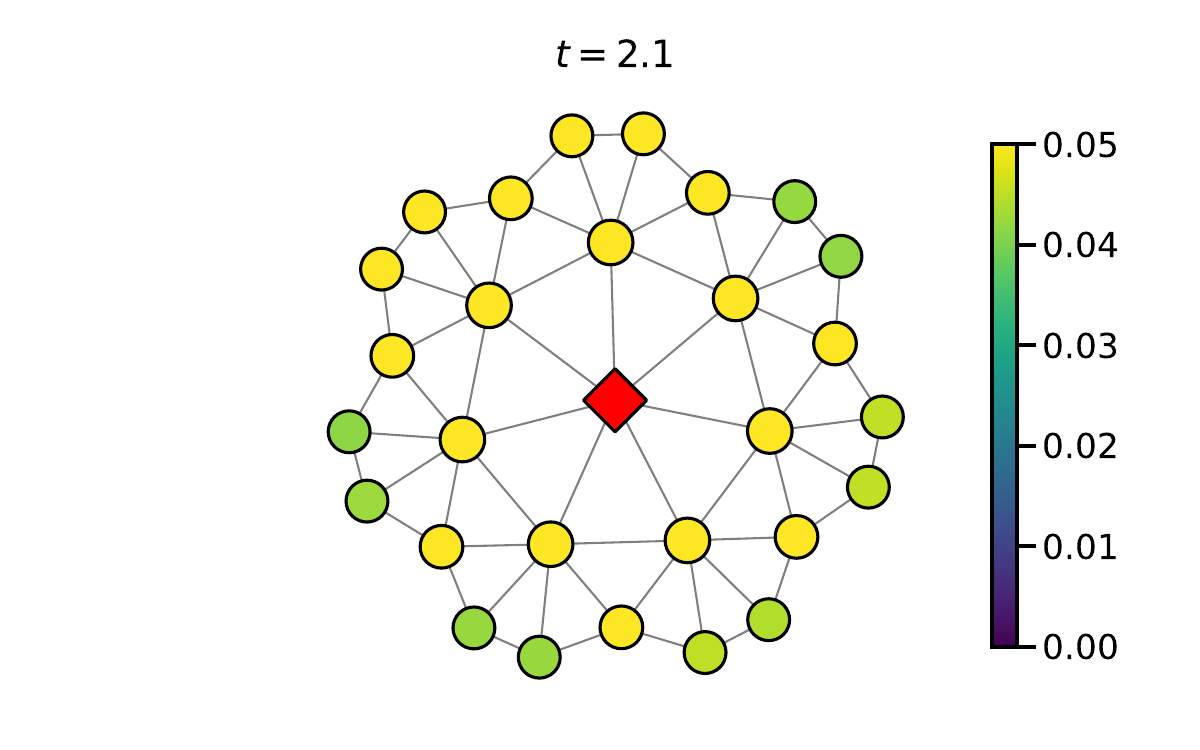}
    }\hfill
    \subfloat[$t = 0.42$, source at first boundary node]{
        \includegraphics[width=0.3\textwidth]{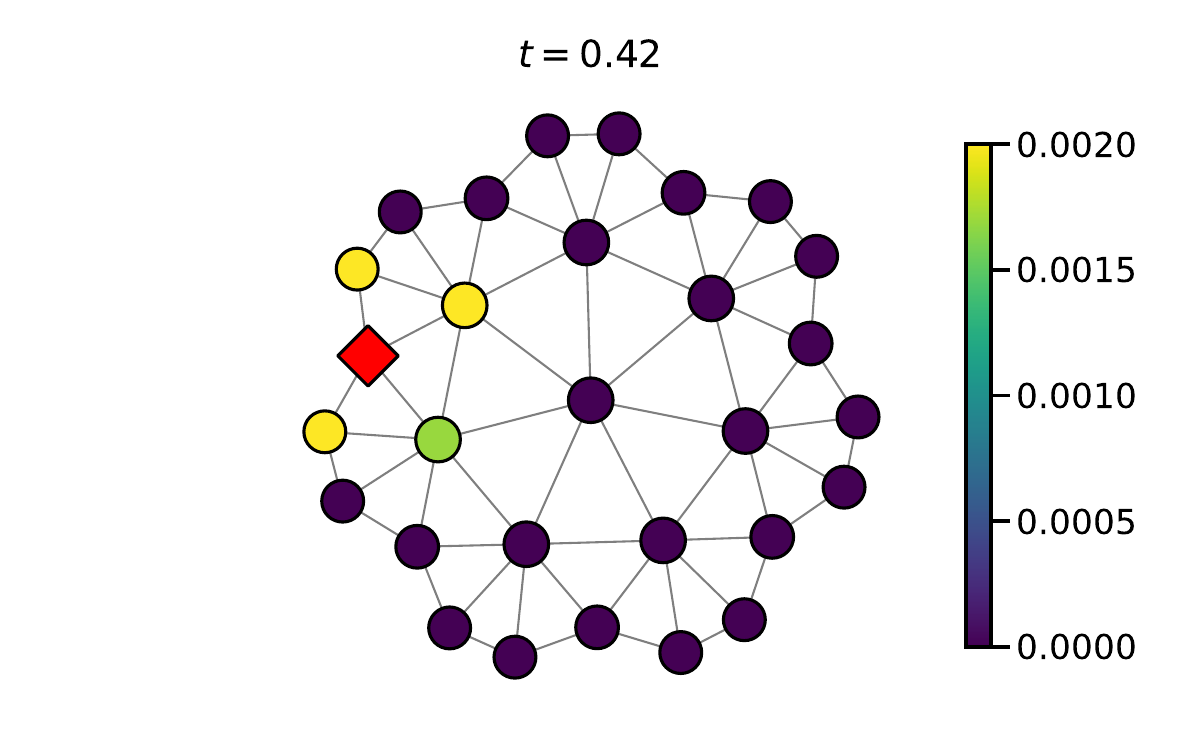}
    }\hfill
    \subfloat[$t = 0.98$, source at first boundary node]{
        \includegraphics[width=0.3\textwidth]{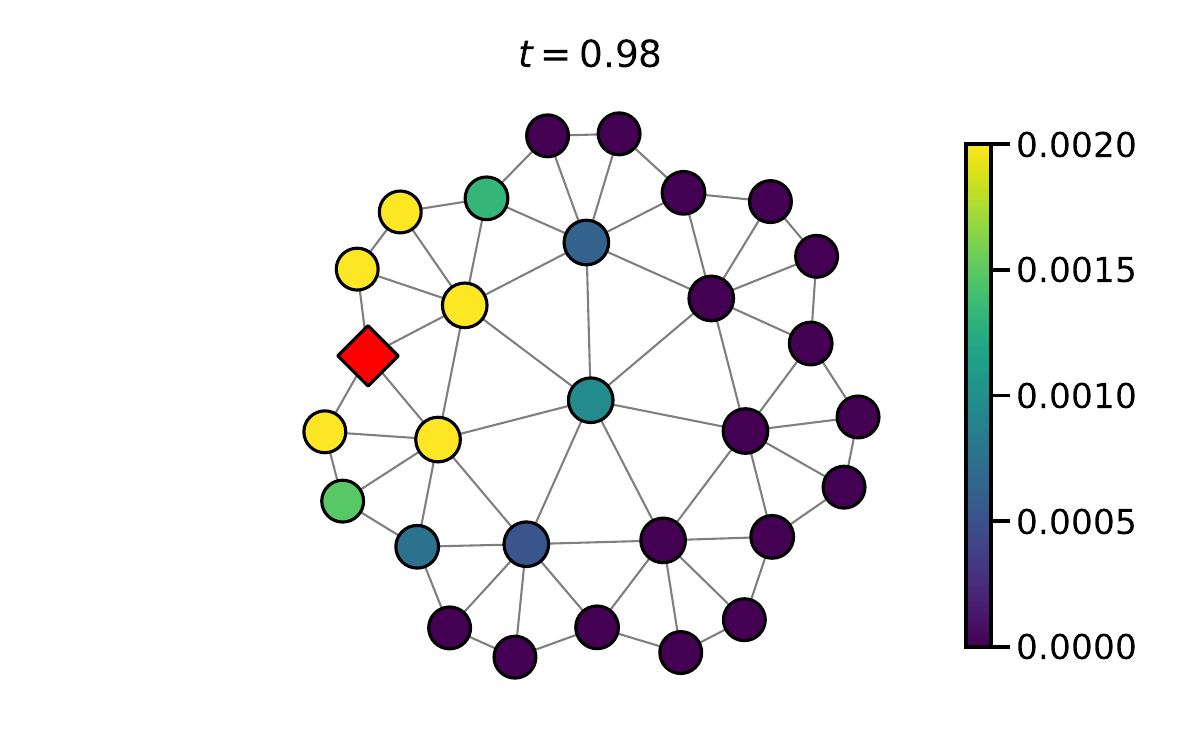}
    }\hfill
    \subfloat[$t = 2.1$, source at first boundary node]{
        \includegraphics[width=0.3\textwidth]{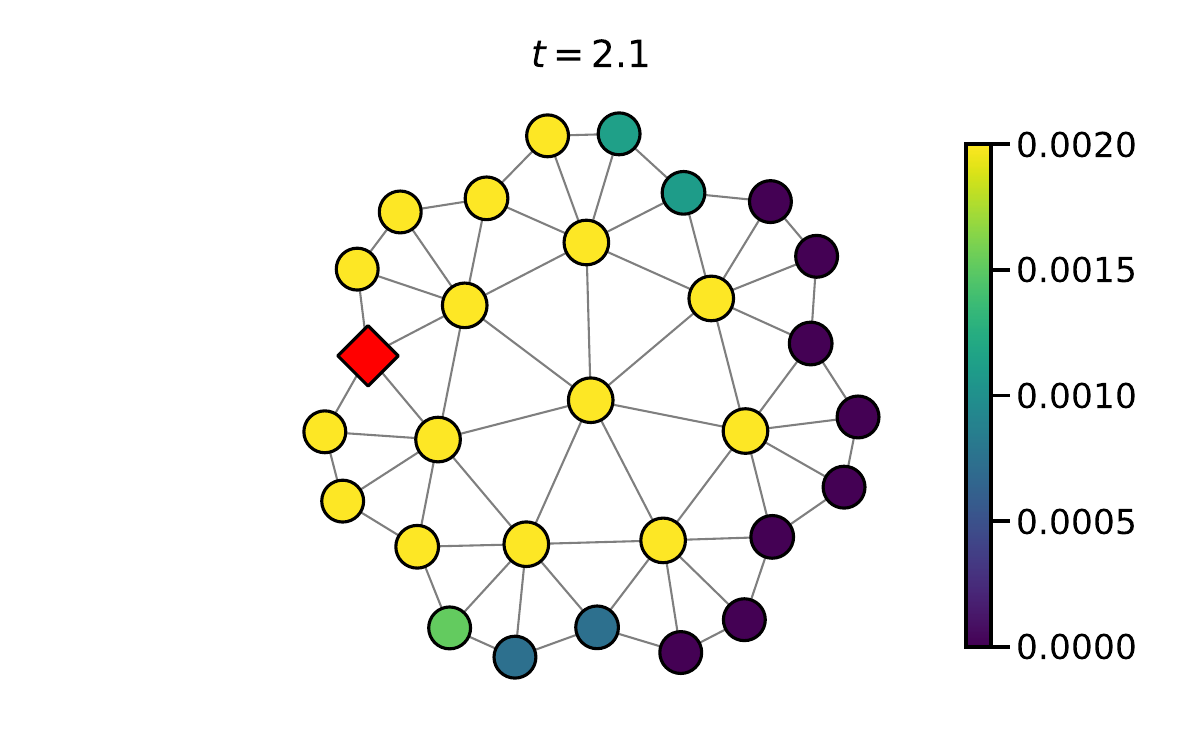}
    }
    \caption{(a),(b),(c) - heatmaps for OTOCs showing information spreading for the case when the center node is the source point while all other nodes act as sink.
    (d),(e),(f) - heatmaps for OTOCs showing information spreading for the case when the first boundary node is the source point while all other nodes act as sink.}
    \label{fig:heatmap}
\end{figure*}

It is also useful to construct heatmap plots that show the magnitude of
the OTOC as a function of position for several successive times. These
are shown in Fig.~\ref{fig:heatmap} (a)-(c) at three different time-slices for an initial source at the center site and Fig.~\ref{fig:heatmap} (d)-(e) when the source is at the first boundary site. The sources are diamond shaped and colored in red. In both cases we observe the spreading of the quantum state as a function of time in agreement with expectations - that is information spreads into non-local degrees of freedom via paths through neighboring spins.

In the case where the source is placed on the boundary, we see that information propagates towards the bulk first and not along the boundary, this is consistent with the geodesics in hyperbolic spaces, where the shortest path between two boundary sites is not on the boundary but goes through the bulk. 
However, 
our MPS ansatz does not exhibit exact rotational symmetry, and this is
reflected in the heatmap plots, Fig.~\ref{fig:heatmap}.

\section{Summary and Conclusion}
\label{sec:conclusions}
In this work we have shown that using a matrix
product state ansatz and a variety of standard tensor network
algorithms, one can study simple Hamiltonian theories residing on tessellations of hyperbolic space. In the continuum
limit, this enables us to probe quantum field theories on three-dimensional anti-de Sitter space. In practice, we
have studied the quantum Ising model as the simplest system in this class and
have restricted our calculations to systems with no more than $232$ spins.
Notice that this is still vastly larger than what could be achieved using
exact diagonalization.

One can think of this lattice system as a simple discrete model for understanding certain aspects of holography. 
To this end, we have computed spatial boundary correlation
functions at fixed time and for different values of the bulk
coupling, finding a power law behavior for any value of the bulk
coupling in the disordered phase, as one expects on the basis
of the (discrete) symmetries of hyperbolic space. Indeed our results
suggest that the boundary theory corresponds to a free Dirac fermion when the bulk is tuned to criticality.  

Our computations of the boundary entanglement entropy however, show that the logarithmic dependence on subregion size expected for a true  CFT is only realized at the bulk critical point - away from
this coupling, a linear scaling is seen. We attribute this to the fact that
the effective CFT gotten by tracing out the bulk degrees of freedom
is generically non-local unless we tune to the bulk critical point. Correspondingly the absence of a bulk graviton suggests the absence of a local energy-momentum tensor in the
boundary CFT.

We have also examined out of time ordered (OTOC) correlators for both bulk and boundary spins showing some evidence of information scrambling.

In general our measurements of
dynamics highlight some of the problems associated with the MPS ansatz: for example we see a lack of rotational invariance in the heatmaps that showcase
time evolution. In addition, the tensor network approach appears to
be limited to modest system sizes of just a few hundred
degrees of freedom. To simulate
larger systems we will need to employ better classical algorithms or quantum computers. 

\begin{acknowledgments}
Numerical computations were performed at Syracuse University HTC Campus Grid under NSF award ACI-1341006. We acknowledge support from U.S. Department of Energy grants DE-SC0019139 and DE-SC0009998. GCT and AFK acknowledge financial support from the National Science Foundation under award No. PHY-2325080: PIF: Software-Tailored Architecture for Quantum Co-Design.
\end{acknowledgments}

\bibliography{refs}

\end{document}